\documentclass[nofootinbib,superscriptaddress,prl,twocolumn,showpacs,preprintnumbers,amsmath,amssymb,bibnotes,altaffilletter]{revtex4-1}
\usepackage{graphicx}
\usepackage{color}

\def\nuc#1#2{${}^{#1}$#2}
\def\mee{$\langle m_{\beta\beta} \rangle$}

\def\BBz{$0\nu\beta\beta$}

\def\Mz{$M_{0\nu}$}

\def\Gz{$G_{0\nu}$}					
\def\Tz{$T^{0\nu}_{1/2}$}

\def\gA{$g_{A}$}                  
\def\qval{$Q_{\beta\beta}$}                 
\def\be{\begin{equation}}
\def\ee{\end{equation}}
\def\cpKkgy{counts/(keV kg yr)}

\def\ppc{P-PC}                          

\def\MJ{{\sc Majorana}}             
\def\DEM{{\sc Demonstrator}}             

\begin{document}
\title{Search for Neutrinoless Double Beta Decay in \nuc{76}{Ge} with the \MJ\ \DEM}

\newcommand{\blhill}{Department of Physics, Black Hills State University, Spearfish, SD 57799, USA}
\newcommand{\ITEP}{National Research Center ``Kurchatov Institute'' Institute for Theoretical and Experimental Physics, Moscow, 117218 Russia}
\newcommand{\JINR}{Joint Institute for Nuclear Research, Dubna, 141980 Russia}
\newcommand{\lbnl}{Nuclear Science Division, Lawrence Berkeley National Laboratory, Berkeley, CA 94720, USA}
\newcommand{\lanl}{Los Alamos National Laboratory, Los Alamos, NM 87545, USA}
\newcommand{\queens}{Department of Physics, Engineering Physics and Astronomy, Queen's University, Kingston, ON K7L 3N6, Canada}
\newcommand{\uw}{Center for Experimental Nuclear Physics and Astrophysics, and Department of Physics, University of Washington, Seattle, WA 98195, USA}
\newcommand{\alberta}{Centre for Particle Physics, University of Alberta, Edmonton, AB T6G 2E1, Canada}
\newcommand{\uchic}{Department of Physics, University of Chicago, Chicago, IL 60637, USA}
\newcommand{\unc}{Department of Physics and Astronomy, University of North Carolina, Chapel Hill, NC 27514, USA}
\newcommand{\ucne}{Department of Nuclear Engineering, University of California, Berkeley, CA 94720, USA} 
\newcommand{\ucph}{Department of Physics, University of California, Berkeley, CA 94720, USA}
\newcommand{\duke}{Department of Physics, Duke University, Durham, NC 27708, USA}
\newcommand{\ncsu}{Department of Physics, North Carolina State University, Raleigh, NC 27695, USA}	
\newcommand{\ornl}{Oak Ridge National Laboratory, Oak Ridge, TN 37830, USA}
\newcommand{\ou}{Research Center for Nuclear Physics, Osaka University, Ibaraki, Osaka 567-0047, Japan}
\newcommand{\pnnl}{Pacific Northwest National Laboratory, Richland, WA 99354, USA}
\newcommand{\princeton}{Department of Physics, Princeton University, Princeton, NJ 08544, USA}
\newcommand{\ttu}{Tennessee Tech University, Cookeville, TN 38505, USA}
\newcommand{\sdsmt}{South Dakota School of Mines and Technology, Rapid City, SD 57701, USA}
\newcommand{\sjtu}{Shanghai Jiao Tong University, Shanghai, China}
\newcommand{\usc}{Department of Physics and Astronomy, University of South Carolina, Columbia, SC 29208, USA}
\newcommand{\usd}{Department of Physics, University of South Dakota, Vermillion, SD 57069, USA} 
\newcommand{\ut}{Department of Physics and Astronomy, University of Tennessee, Knoxville, TN 37916, USA}
\newcommand{\tunl}{Triangle Universities Nuclear Laboratory, Durham, NC 27708, USA}
\newcommand{\mpi}{Max-Planck-Institut f\"{u}r Physik, M\"{u}nchen, 80805 Germany}
\newcommand{\tum}{Physik Department and Excellence Cluster Universe, Technische Universit\"{a}t, M\"{u}nchen, 85748 Germany}

\affiliation{\pnnl}
\affiliation{\lbnl}
\affiliation{\uw}
\affiliation{\usc}
\affiliation{\ornl}
\affiliation{\ncsu}
\affiliation{\tunl}
\affiliation{\ITEP}
\affiliation{\uchic}
\affiliation{\usd}
\affiliation{\mpi}
\affiliation{\sdsmt}
\affiliation{\lanl}
\affiliation{\unc}
\affiliation{\JINR}
\affiliation{\duke}
\affiliation{\ut}
\affiliation{\ou}
\affiliation{\princeton}
\affiliation{\alberta}
\affiliation{\ttu}
\affiliation{\ucph}
\affiliation{\sjtu}
\affiliation{\queens} 
\affiliation{\tum}
\affiliation{\blhill}
\affiliation{\ucne}

 \author{C.E.~Aalseth}\affiliation{\pnnl}
\author{N.~Abgrall}\affiliation{\lbnl}		
\author{E.~Aguayo}\altaffiliation{Present address: Intel Corp., 2111 NE 25th Ave, Hilsboro, OR 97124, USA}\affiliation{\pnnl}
\author{S.I.~Alvis}\affiliation{\uw}	
\author {M.~Amman}\affiliation{\lbnl}
\author{I.J.~Arnquist}\affiliation{\pnnl} 
\author{F.T.~Avignone~III}\affiliation{\usc}\affiliation{\ornl}
\author{H.O.~Back}\altaffiliation{Present address: \pnnl}\affiliation{\ncsu}\affiliation{\tunl} 
\author{A.S.~Barabash}\affiliation{\ITEP}
\author{P.S.~Barbeau}\altaffiliation{Present address: \duke, \tunl}\affiliation{\uchic}  
% ...
			
\collaboration{{\sc{Majorana}} Collaboration}
\noaffiliation

\begin{abstract}
The \MJ\ Collaboration is operating an array of high purity Ge detectors to search for neutrinoless double-beta decay in $^{76}$Ge. The \MJ\ \DEM\  comprises 44.1~kg of Ge detectors (29.7 kg enriched in $^{76}$Ge) split between two modules contained in a low background shield at the Sanford Underground Research Facility in Lead, South Dakota. Here we present results from data taken during construction, commissioning, and the start of full operations. We achieve unprecedented energy resolution of 2.5 keV FWHM at \qval\  and a very low background with no observed candidate events in 9.95 kg yr of enriched Ge exposure, resulting in a lower limit on the half-life of $1.9\times10^{25}$ yr (90\% CL). This result constrains the effective Majorana neutrino mass to below 240 to 520 meV, depending on the matrix elements used. In our experimental configuration with the lowest background, the background is $4.0_{-2.5}^{+3.1}$ counts/(FWHM t yr).
 \end{abstract}


\maketitle

Searches for neutrinoless double-beta (\BBz) decay test the Majorana nature of the neutrino~\cite{Zralek1997}. The observation of this process would imply that total lepton number is violated and that neutrinos are Majorana particles~\cite{sch82}. A measurement of the \BBz\ decay rate may also yield information on the absolute neutrino mass. Measurements of atmospheric, solar, and reactor neutrino oscillation~\cite{PDB16} indicate a large parameter space for the discovery of \BBz\ decay. Moreover, evidence from the SNO experiment~\cite{Ahm04} of a clear departure from maximal mixing in solar neutrino oscillation implies a minimum \mee\ of $\sim$15 meV for the inverted mass ordering scenario. This target is within reach of next-generation \BBz\ searches. An experiment capable of observing this minimum rate would therefore help elucidate the Majorana or Dirac nature of the neutrino for inverted-ordering neutrino masses. Even if the ordering is normal, these experiments will have very high discovery probability~\cite{Agostini2017a}; a null result would improve the existing sensitivity by $\sim$1 order of magnitude. A nearly background-free tonne-scale \nuc{76}{Ge} experiment would be sensitive to effective Majorana neutrino masses below $\sim$15~meV. For recent comprehensive experimental and theoretical reviews on \BBz, see Refs.~\cite{avi08, bar11, Rode11, Elliott2012, Vergados2012, Cremonesi2013, Schwingenheuer2013, Elliott2015, Henning2016}.

The \MJ\ \DEM\ is an array of isotopically enriched and natural Ge detectors searching for the decay of isotope \nuc{76}{Ge}. A primary technical goal of the experiment is the development and use of ultra-low activity materials and methods to suppress backgrounds to a low enough level to motivate the construction of a tonne-scale experiment.  Here we present first results on the achieved background level and the corresponding \BBz\ limit from an analysis of an initial detector exposure of 9.95 kg yr.

The \DEM\ utilizes the well-known benefits of enriched high-purity germanium (HPGe) detectors, including intrinsically low-background source material, understood enrichment process, excellent energy resolution, and sophisticated event reconstruction.  We have assembled two modular HPGe arrays fabricated from ultra-pure electroformed copper. 
The enriched detectors are P-type, point-contact (\ppc) HPGe detectors~\cite{luk89,Barbeau2007,Aguayo2011}. These detectors allow a low-energy threshold permitting a variety of physics studies~\cite{Abgrall2017a}.

Each of the \DEM\ detectors has a mass of 0.6-1.1 kg. The two cryostats contain 35 detectors with a total mass of 29.7 kg fabricated with Ge material enriched to 88.1$\pm$0.7\% in \nuc{76}{Ge} as measured by ICP-MS, and 23 detectors with a total mass of 14.4 kg fabricated from natural Ge (7.8\% \nuc{76}{Ge}).  The 69.8\% yield of converting initial enriched material into Ge diodes is the highest achieved to date~\cite{Abgrall2017c}. Module~1(2) houses 16.8~kg (12.9~kg) of enriched germanium detectors and 5.6~kg (8.8~kg) of natural germanium detectors. The two modules were installed sequentially with data collected from Module 1 (M1) while Module 2 (M2) was assembled.

A detailed description of the experimental setup can be found in Ref.~\cite{Abgrall2014} and some initial results were reported in Ref.~\cite{Elliott2016}. 
Starting from the innermost cavity, two cryostat modules are surrounded by an inner layer of electroformed copper, an outer layer of commercially obtained C10100 copper, high-purity lead, an active muon veto~\cite{Bugg2014}, borated polyethylene, and polyethylene. The cryostats, copper, and lead shielding are all enclosed in a radon exclusion box that is purged with liquid-nitrogen boil-off gas. The experiment is located in a clean room at the 4850-foot level (1478 m, 4300 m.w.e.) of the Sanford Underground Research Facility (SURF) in Lead, South Dakota~\cite{Heise2015}. The radioassay program developed to ensure the apparatus met background goals is described in Ref.~\cite{Abgrall2016}. A parts-tracking database used to monitor exposures and inventory control is described in Ref.~\cite{Abgrall2015}. High voltage testing of components is described in Ref.~\cite{Abgrall2016a} and the low-background readout electronics are described in Ref.~\cite{Barton2011,Guinn2015}.  

\begin{figure}[h]
 \centering
 \includegraphics[width=0.95\columnwidth, keepaspectratio=true]{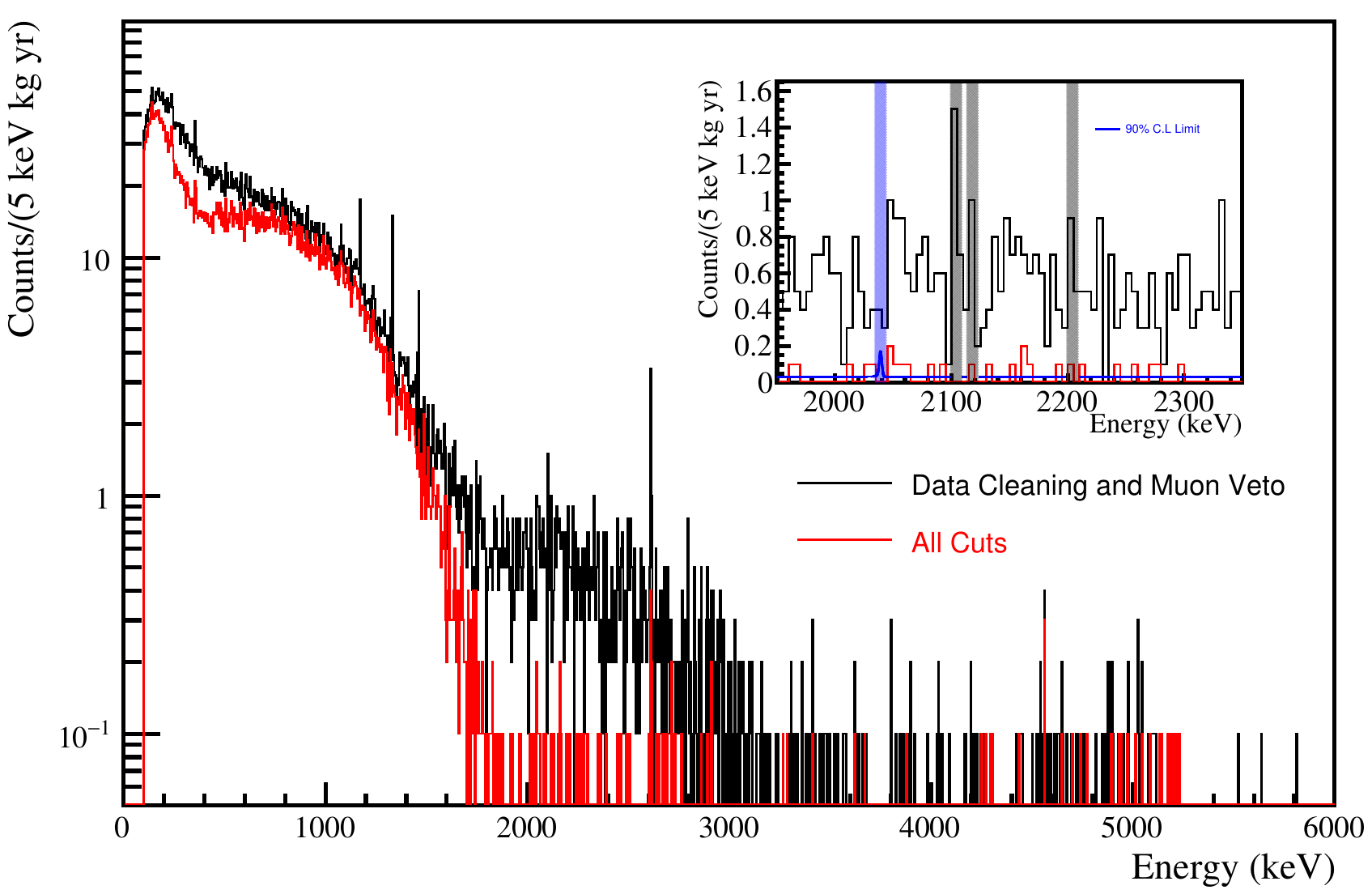}
 \includegraphics[width=0.95\columnwidth, keepaspectratio=true]{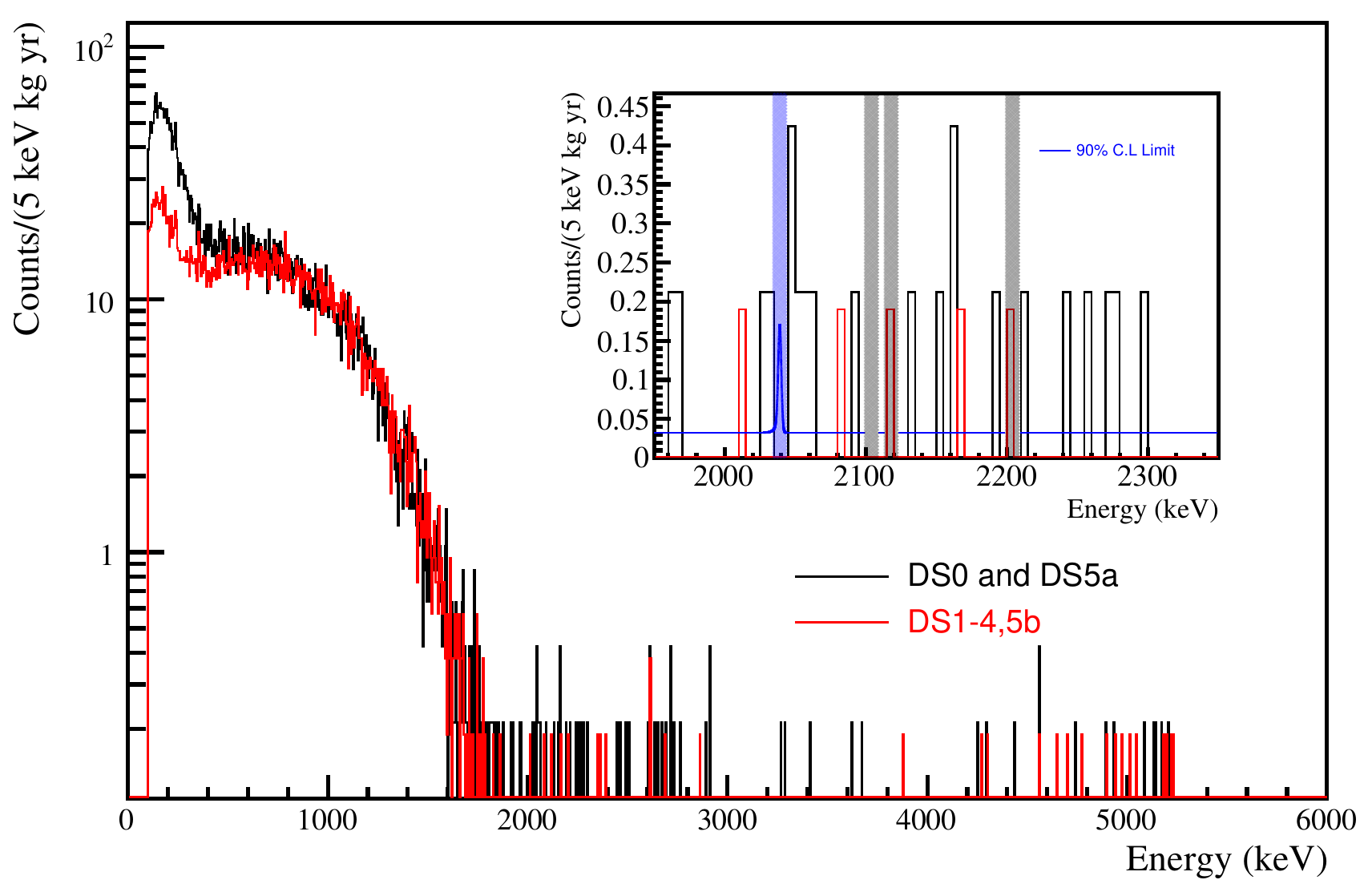}
 \caption{Color online. Top: The spectrum above 100 keV of all six data sets summed together with only data reduction and muon veto cuts (Black) and after all cuts (Red).  Bottom: The spectrum above 100 keV after all cuts from the higher background data sets DS0 and DS5a (Black) compared to the data sets with lower background, DS1-4,5b (Red).  Note the $\gamma$ background in DS0 is higher and the $\alpha$ rate is the same without pulse shape analysis. Rejection of $\alpha$s in DS5a is degraded due to noise as described in the text. Insets: The same as in the primary plots but for the 360-keV region. The blue and graded shaded regions are excluded when determining the background. The thin blue curves shows the 90\% CL upper limit for \BBz\ at \qval\ as described in the text, which corresponds to 2.04 signal counts. }
 \label{Fig:Background}
\end{figure}

The data presented here are subdivided into six data sets, referred to as DS0 through DS5, distinguished by significant experimental configuration changes. DS0 was a set of commissioning runs and was terminated to install the inner 2-inch electroformed copper shield and additional shielding. As a result, DS1 showed significantly reduced background. DS1 was terminated in order to test multisampling of the digitized waveforms, providing extended signal capture following an event for improved alpha background rejection. DS2 was terminated for the installation of Module 2.  DS3 and DS4 consist of data taken from Module 1 and Module 2, respectively, with separate DAQ systems. DS5 consists of data taken after the DAQ systems were merged. The final installations of poly shielding enclosing the apparatus extended into DS5. We thus subdivided DS5 into two sub-ranges, DS5a and DS5b, where the latter corresponds to data taken after the detector was fully enclosed within the initial layer of poly shielding, allowing the establishment of a robust grounding scheme that reduced the electronic noise. The noise in DS5a impacted the pulse shape analysis, resulting in degraded background rejection.  These changes define the difference between the data sets, with the combination of DS1-4 and 5b having the lowest background (see the lower panel in Fig.~\ref{Fig:Background}).

The detector is calibrated using periodic ($\sim$weekly) \nuc{228}{Th} line source calibration runs~\cite{Abgrall2017}. Event energies are reconstructed from the pulse amplitudes, using a trapezoidal filter algorithm whose parameters are tuned to minimize calibration source gamma line widths. We correct for (hole) charge trapping using the measured charge drift times, which greatly improves the resolution. The parameters of the peak shape are fit as a function of energy, which at the \BBz\ {\it Q} value (\qval) yields a mixture ($\gtrsim$4:1) of a Gaussian ($\sigma \sim$1 keV) plus exGaussian (same $\sigma$, $\tau \sim$2 keV), where the parameters are determined individually for each data set. This peak shape yields an average FWHM at \qval\ of 2.52$\pm$0.08 keV (excluding DS5a), the best achieved to-date for a neutrinoless double-beta decay search. The uncertainty accounts for time variation, residual ADC non-linearities, and statistical uncertainties. Including DS5a, the average resolution is 2.66$\pm$0.08 keV.

Table~\ref{tab:DataSets} summarizes the key features of each data set. During each set, some detectors were inoperable. Furthermore, the system was not always collecting physics data due to calibration, systematic checks and construction activities. The live time and the active mass numbers within the table reflect these conditions.

The active mass calculations take into account the detector dead layers, measured with collimated \nuc{133}{Ba} source scans, as well as the detailed shape of each detector measured with an optical scanner. The active mass is $\sim$90\% of the total mass with a systematic uncertainty of $\sim$1.4\%, which dominates the uncertainty in the exposure. The exposure calculation, performed detector-by-detector, accounts for periods in which detectors are temporarily removed from the analysis due to instabilities. 
The exposure includes corrections for dead-time due to a 15-min cut to remove microphonic events coincident with liquid nitrogen fills and a 1-s period following muons detected in the veto system~\cite{Abgrall2017b}. It also includes small losses due to the time for the digitizers to re-trigger after an event and cuts to remove pulser events, which are used to estimate dead time losses, gain shifts, and other instabilities. These cuts reduce the total live time by 1-5\% depending on data set. The uncertainty in the live time is $<$0.5\%. The results presented here are based on our open data. The blinded fraction of our data, set by a data parsing scheme, will be presented in future publications.

Data from the \DEM\ are first filtered by data cleaning routines to remove non-physical waveforms while retaining $>$99.9\% of true physics events ($\epsilon_{DC}$). Double-beta decay events are characterized as single-site events because the electrons deposit their energy over a small range ($\sim$1~mm) compared to our ability to distinguish separate charge deposition sites as would arise from a typical multiple-Compton-scattered background $\gamma$. We reject any events that trigger more than one detector within a 4-$\mu$s time window; the small associated signal inefficiency is negligible but we account for the associated dead time in our exposure calculation. Next we remove events whose waveforms are typical of multi-site energy deposits ($\epsilon_{AE}$) while retaining 90\%$\pm$3.5\% of single-site events. This pulse shape discrimination~\cite{Cuesta2016a} is based on the relationship between the maximum current and energy, similar to the cut described in Ref.~\cite{Agostini2013c}, and is similarly tuned using \nuc{228}{Th} calibration source data. The efficiency uncertainty accounts for channel-, energy- and time-variation, as well as for the position distribution difference between calibration and \BBz\ events, established using the MaGe simulation framework~\cite{Bos11} built against Geant4~\cite{Agostinelli2003250} and the detector signal simulation package, siggen~\cite{RadfordSiggen}.

The analysis also removes events that arise from external $\alpha$ particles impinging upon the passivated surface of our \ppc\ detectors.   For such events, electron drift is negligible, and significant charge trapping of holes occurs in the immediate vicinity of the surface. The reconstructed energy of these events corresponds to the fraction of energy collected within the shaping time of our energy filter, resulting in energy degradation that sometimes populates the energy region near \qval. However, subsequent release of the trapped charge results in a significantly increased slope in the tail of the signal pulses. Quantification of this delayed charge recovery permits a highly effective reduction of this potential background using pulse shape discrimination~\cite{Gruszko2016} while retaining 99$\pm$0.5\% ($\epsilon_{DCR}$) of the bulk-detector events, as measured using \nuc{228}{Th} calibration source events. Collimated $\alpha$-source measurements with a \ppc\ detector show that the waveform response and energy spectrum are a strong function of the distance from the point contact where the $\alpha$ impinges upon the passivated surface. The results indicate that the activity is most likely due to \nuc{210}{Pb}-supported \nuc{210}{Po} plated out as Rn daughters on the Teflon components of the detector mount. 

True \BBz\ events can exhibit energy degradation far from the \qval\ due to their proximity to the crystal dead layer or due to the emission of bremsstrahlung resulting in energy deposition not fully contained within the active detector volume.  Using the MaGe simulation framework, we estimate that 91$\pm$1\% of true \BBz\ events are fully contained ($\epsilon_{cont}$). The uncertainty accounts for uncertainties in the detector geometry and the difference between simulation and literature values for bremsstrahlung rates and electron range.

All efficiencies are calculated individually for each data set with values listed in Table~\ref{tab:DataSets}. The product of the number of \nuc{76}{Ge} atoms ($N$), the live time ($T$), and the total signal efficiency ($\epsilon_{tot}$) for each data set are also summarized in Table~\ref{tab:DataSets}\footnote{GERDA treats the live time fraction, isotopic fraction and detector active volume as efficiencies~\cite{Agostini2017}. \MJ\ treats those as atomic exposure factors. }.

\begin{table*}[htp]
\caption{A summary of the key parameters of each data set. The exposure calculation is done detector-by-detector. Symmetric uncertainties for the last digits are given in parentheses. The exposure numbers are for open data. }
\begin{center}
\begin{tabular}{lclccccccc}
\hline
Data 	&	Start 		& Hardware 		&Active Enr. 			& Exposure 			& $\epsilon_{AE}$	& $\epsilon_{DCR}$	& $\epsilon_{cont}$	& $\epsilon_{tot}$	& 	$N T \epsilon_{tot} \epsilon_{res}$  	\\
Set		&	Date			&Distinction		&	Mass (kg)				&(kg-yr)			&				&				&				&					& 	($10^{24}$atoms y)\\
\hline\hline
DS0		&	6/26/15	& No Inner Cu Shield	& 10.69(16)			&	\phantom01.26(02)	&0.901$_{-0.035}^{+0.032}$&0.989$_{-0.002}^{+0.009}$&0.908(11)		&0.808$_{-0.033}^{+0.031}$	&\phantom0$6.34_{-0.27}^{+0.25}$		\\	
DS1		&	12/31/15	& Inner Cu Shield added	& 11.90(17)			&	\phantom01.81(03)	&0.901$_{-0.040}^{+0.036}$&0.991$_{-0.005}^{+0.010}$&0.909(11)		&0.811$_{-0.038}^{+0.035}$	&\phantom0$9.23_{-0.44}^{+0.41}$			\\
DS2		&	5/24/16	& Multisampling		& 11.31(16)			&	\phantom00.29(01)	&0.903$_{-0.037}^{+0.035}$&0.986$_{-0.005}^{+0.011}$&0.909(11)		&0.809$_{-0.035}^{+0.034}$	&\phantom0$1.49_{-0.07}^{+0.06}$			\\
DS3		&	8/25/16	& M1 and M2 installed	& 12.63(19)			&	\phantom01.01(01)	&0.900$_{-0.031}^{+0.030}$&0.990$_{-0.003}^{+0.010}$&0.909(11)		&0.809$_{-0.030}^{+0.030}$	&\phantom0$5.18_{-0.20}^{+0.20}$		\\	
DS4		&	8/25/16	& M1 and M2 installed	&  \phantom05.47(08)	&	\phantom00.28(00)	&0.900$_{-0.034}^{+0.031}$&0.992$_{-0.002}^{+0.011}$&0.908(10)		&0.809$_{-0.032}^{+0.030}$	&\phantom0$1.47_{-0.06}^{+0.06}$			\\
DS5a	&	10/13/16	& Integrated DAQ (noise)	& 17.48(25)			&	\phantom03.45(05)	&0.900$_{-0.036}^{+0.034}$&0.969$_{-0.013}^{+0.013}$&0.909(13)		&0.792$_{-0.035}^{+0.034}$	&$17.17_{-0.79}^{+0.76}$					\\	
DS5b	&	1/27/17	& Optimized Grounding	& 18.44(26)			&	\phantom01.85(03)	&0.900$_{-0.033}^{+0.031}$&0.985$_{-0.005}^{+0.014}$&0.909(13)		&0.805$_{-0.032}^{+0.032}$	&\phantom0$9.46_{-0.39}^{+0.39}$					\\
\hline
Total		&				&				&						&	9.95(21)			&					&				&				&			&$50.35_{-0.73}^{+0.70}$					\\
Total		&	(DS1-4,5b)	&				&						&	5.24(17)			&					&				&				&			&$26.84_{-0.68}^{+0.65}$						\\
\hline\hline
\end{tabular}
\end{center}
\label{tab:DataSets}
\end{table*}%

Figure~\ref{Fig:Background} shows the measured event spectrum above 100~keV using our event selection criteria. Background projections based on our assay program and MaGe simulations predict a flat background between 1950 keV and 2350 keV after rejecting possible gamma peaks within that energy range. We exclude $\pm$5 keV ranges (indicated in the figure by shading) centered at 2103 keV (\nuc{208}{Tl} single escape peak), 2118 keV and 2204 keV (\nuc{214}{Bi}).  We also ignore events near \qval\ between 2034 keV and 2044 keV. The backgrounds are summarized in Table~\ref{tab:Background} along with the corresponding continuum background index (BI) assuming a flat profile.  Uncertainties are computed using Feldman-Cousins intervals~\cite{Feldman1998}. All six data sets are nearly background free at the present exposure and therefore we use all for a \Tz\ limit.

\begin{table}[htp]
\caption{The background (BG) within the defined 360-keV window for each of the data sets. The background index (BI, column 3) is given in units of counts/(keV kg yr).  The optimum ROI width for each data set is given in column 4 and the resulting expected background counts within that ROI is given in the final column. The final row provides a summary for the lowest expected background data partition.}
\begin{center}
\begin{tabular}{lcrcccc}
\hline
Data 		&	Window			& BI	 						&ROI 			& ROI BG 		\\
Set			&	Counts			&$\times10^{-3}$				& (keV)			& (counts)			\\
\hline\hline
DS0			&	11				& 24.3$_{-7.0}^{+8.4}$			& 3.93			&	0.120				\\	
DS1			&	3				& \phantom04.6$_{-2.9}^{+3.5}$	& 4.21			&	0.035					\\	
DS2			&	0				& $<$12.3						& 4.34			&	0.000							\\	
DS3			&	0				& $<$3.6						& 4.39			&	0.000						\\	
DS4			&	0				&$<$12.7						& 4.25			&	0.000							\\	
DS5a		&	10				& \phantom08.0$_{-2.6}^{+3.1}$	& 4.49			&	0.125					\\	
DS5b		&	0				& $<$1.9						& 4.33			&	0.000					\\	
\hline
Total			&	24				&	\phantom06.7$_{-1.4}^{+1.4}$	& 4.32			&	0.288							\\
DS1-4,5b		&	3				&	\phantom01.6$_{-1.0}^{+1.2}$	& 				&	0.036						\\
\hline\hline
\end{tabular}
\end{center}
\label{tab:Background}
\end{table}%

After all cuts in DS0-5, there are 24 events within the 360-keV background window. This results in a background index normalized to active mass for the summed spectrum of 17.8$\pm$3.6 counts/(FWHM t yr) or $(6.7\pm1.4)\times10^{-3}$~\cpKkgy. In the context of background rates relevant for future data taking and for a next-generation \BBz\ experiment, we compute  the background from the lowest expected background configuration (DS1-4,5b), which corresponds to $4.0_{-2.5}^{+3.1}$ counts/(FWHM t yr) or (1.6$_{-1.0}^{+1.2})\times10^{-3}$~\cpKkgy. This background level is statistically consistent with GERDA's best-achieved background to date for a \BBz\ search, $2.9^{+1.8}_{-1.2}$ counts/(FWHM t y) for a resolution of 2.9 keV~\cite{Pandola2017,Agostini2018}.

The half-life limit can be approximated as a Poisson-process search in an optimized region-of-interest (ROI) surrounding the peak energy. The ROI is optimized based on the achieved background level similarly to Ref.~\cite{Agostini2017a}  except that the measured peak-shape function was used in place of the Gaussian assumption\footnote{See Eqn.~B4 in Ref.~\cite{Agostini2017a}.}. The result varies for individually-considered data sets as given in Table~\ref{tab:Background} and the exposure-weighted-average optimal ROI is 4.32 keV, with corresponding additional efficiency $\epsilon_{res}$ = 0.899$\pm$0.005.  This resolution efficiency factor only applies for this counting-measurement analysis and not for the spectrum fits described later. The measured background index corresponds to 0.29 expected total background counts in the optimized ROI near \qval. The lower limit on the half-life is thus approximately given by:
\begin{equation}
\mbox{\Tz}  > \frac{\ln(2) N T \epsilon_{tot}\epsilon_{res}}{S},
\label{eq:halflife}
\end{equation}
where $S$ is the upper limit on the number of signal events that can be attributed to \BBz.  Using the Feldman-Cousins approach~\cite{Feldman1998} for 0 observed counts with an expected background of 0.29 gives $S$=2.15 at 90\% CL. This results in a \nuc{76}{Ge} \BBz\ half-life limit of $1.6\times10^{25}$ yr.

We derive our quoted limit using an unbinned, extended profile likelihood method implemented in RooStats~\cite{Verkerke2003,Schott2012,Agostini2017}. The analysis was carried out in an analysis energy range from 1950 to 2350 keV, removing the three 10-keV sub-ranges that correspond to the known $\gamma$ lines predicted by our background projections. The background in the analysis range is modeled as a flat distribution and the \BBz\ signal is modeled using the full peak shape with data-set-specific parameters evaluated as described above. While the hypothetical \BBz\ half-life is universal in all data sets, the exposures, the peak shape parameters, and the analysis cut efficiencies are constrained near their dataset-specific values using Gaussian nuisance terms in the likelihood function. Monte Carlo simulations were performed for the Neyman interval construction. The p-value distribution for this method finds the observed lower limit on the \BBz\ decay half life is $1.9\times10^{25}$ yr at 90\% CL. The 90\% CL median sensitivity for exclusion is $>$$2.1\times10^{25}$~yr. We chose to quote this result because it has reliable coverage by construction, based on simulations. GERDA also follows this approach, which facilitates comparison.

We explored several alternative statistical analyses of our data.  A modified profile likelihood that examines the ratio of the p-values of the background-plus-signal model and the background-only model (the CL$_{\rm s}$ method~\cite{Read2000}) mitigates the effect of background down-fluctuations. The CL${_{\rm s}}$ method yields  $1.5 \times10^{25}$ yr as the observed limit and $1.4\times10^{25}$ yr as the median sensitivity on the half life at 90\% confidence level. Additionally, a Bayesian analysis was carried out with Markov chain Monte Carlo simulations also using the RooStats software and the same likelihood function as our primary analysis. Assuming a flat prior on $\Gamma \equiv 1/T^{0\nu}_{1/2}$, the Bayesian limit on the half life is $1.6 \times10^{25}$ yr for a 90\% credible interval. Using instead the Poisson Jeffreys prior flat in $\sqrt{\Gamma}$ yields a limit of $2.6\times10^{25}$ yr. 

To place limits on \mee, matrix element (\Mz) and phase space (\Gz) calculations are required. The review in Ref.~\cite{Engel2016} provides an overview of matrix-element theory. Here we use Refs.~\cite{Men09,Horoi2016,Barea2015,Hyvarinen2015,Simkovic2013,Vaquero2013,Yao2015} for an overall range for \Mz\ of 2.81 to 6.13. With this range of \Mz, our limit of \Tz\ = $1.9\times10^{25}$~yr results in \mee\ $<$ 240 to 520 meV, using the \Gz\ of $2.36\times10^{-15}$~/yr~\cite{Kotila2012} or $2.37\times10^{-15}$~/yr~\cite{Mirea2015} and a value of \gA=1.27. 

Despite the presently low exposure, the \DEM\ is approaching limits comparable to the best efforts to date. In fact, the two Ge \BBz-experiments \MJ\ and GERDA~\cite{Agostini2017,Agostini2018} have modest exposures compared to  KamLAND-Zen~\cite{Gando2016} and EXO-200~\cite{Albert2017}. However the very low background and excellent energy resolution help overcome the exposure limitation. Both \MJ\ and GERDA are presently operating in the nearly background free regime. Hence, a combination of results can be approximated by adding half-lives. The present GERDA limit is $8.0\times10^{25}$~yr~\cite{Pandola2017}. A combined limit would therefore be near $10^{26}$~yr. {Selecting the best technologies of these two experiments with comparable backgrounds and excellent resolutions from very distinct configurations indicate that a future larger experiment using \nuc{76}{Ge}, such as LEGEND~\cite{Abgrall2017d},  is warranted.

In summary, the goal of the \MJ\ \DEM\ is to show that backgrounds can be reduced to a value low enough to justify a large \BBz\ experiment using \nuc{76}{Ge}. We have built two modules of HPGe arrays from ultra-low-background components. Initial results indicate the background level is very low. 
The \DEM\ goal is to reach a background of 2.5 counts/(FWHM t yr) and the presented result is consistent with that goal, demonstrating the success of the assay program and other production radioactivity controls. 
At the present level of background, the limit on \Tz\ is increasing nearly linearly and is projected to approach $10^{26}$ yr for a 100 kg yr exposure. 

The authors appreciate the technical assistance of J.F. Amsbaugh, 
J. Bell, 
B.A. Bos, 
T.H. Burritt, 
G. Capps, 
K. Carney, 
R. Daniels,
L. DeBraeckeleer, 
C. Dunagan, 
G.C. Harper, 
C. Havener, 
G. Holman, 
R. Hughes, 
K. Jeskie, 
K. Lagergren, 
D. Lee, 
M. Middlebook, 
A. Montoya, 
A.W. Myers, 
D.Peterson, 
D. Reid, 
L. Rodriguez, 
H. Salazar, 
A.R. Smith, 
G. Swift, 
M. Turqueti, 
J. Thompson, 
P. Thompson, 
C. Tysor, 
T.D. Van Wechel, and
R. Witharm.

This material is based upon work supported by the U.S. Department of Energy, Office of Science, Office of Nuclear Physics under Award  Numbers DE-AC02-05CH11231, DE-AC52-06NA25396, DE-FG02-97ER41041, DE-FG02-97ER41033, DE-FG02-97ER41042, DE-SC0012612, DE-FG02-10ER41715, DE-SC0010254, and DE-FG02-97ER41020. This material is based upon work supported by the National Science Foundation under Grant Nos. MRI-0923142, PHY-0855314, PHY-0919270, PHY-1003399, PHY-1003940, PHY-1102292, PHY-1202950, PHY-1206314, PHY-1619611, and PHY-7194299. We acknowledge support from the Russian Foundation for Basic Research, grant No. 15-02-02919. We  acknowledge the support of the U.S. Department of Energy through the LANL/LDRD Program. This research used resources of the Oak Ridge Leadership Computing Facility, which is a DOE Office of Science User Facility supported under Contract DE-AC05-00OR22725. This research used resources of the National Energy Research Scientific Computing Center, a DOE Office of Science User Facility supported under Contract No. DE-AC02-05CH11231. We thank our hosts and colleagues at the Sanford Underground Research Facility for their support.

\bibliographystyle{iopart-num}
\bibliography{DoubleBetaDecay.bbl}

\end{document}